\documentclass{aa}

\usepackage{graphicx,dblfloatfix,fixltx2e}
\usepackage{txfonts}
\usepackage{hyperref}

\begin{document}

   \title{Intensity contrast of the average supergranule}

   \author{J. Langfellner \inst{1}
             \and
             A.~C. Birch \inst{1}
             \and
             L. Gizon \inst{1,2,3}}

   \institute{Max-Planck-Institut f\"ur Sonnensystemforschung,
             Justus-von-Liebig-Weg 3, 37077 G\"ottingen, Germany
             \and
             Georg-August-Universit\"at, Institut f\"ur Astrophysik,
               Friedrich-Hund-Platz 1, 37077 G\"ottingen, Germany
             \and
             Center for Space Science, NYUAD Institute, New York University Abu
             Dhabi, PO Box 129188, Abu Dhabi, UAE
             }

  \date{Received <date> / Accepted <date>}

  \abstract
   {While the velocity fluctuations of supergranulation dominate the spectrum of solar convection at the solar surface, very little is known about the fluctuations in other physical quantities like temperature or density at supergranulation scale. 
Using SDO/HMI observations, we characterize the intensity contrast of solar supergranulation at the solar surface. 
We identify the positions of ${\sim}10^4$ outflow and inflow regions at supergranulation scales, from which we construct average flow maps and co-aligned intensity and magnetic field maps.
In the average outflow center, the maximum intensity contrast is $(7.8\pm0.6)\times10^{-4}$ (there is no corresponding feature in the line-of-sight magnetic field). This corresponds to a temperature perturbation of about $1.1\pm0.1$~K, in agreement with previous studies. We discover an east-west anisotropy, with a slightly deeper intensity minimum east of the outflow center. The evolution is asymmetric in time: the intensity excess is larger 8~hours before the reference time (the time of maximum outflow), while it has almost disappeared 8~hours after the reference time. In the average inflow region, the intensity contrast mostly follows the magnetic field distribution, except for an east-west anisotropic component that dominates 8~hours before the reference time.
We suggest that the east-west anisotropy in the intensity is related to the wave-like properties of supergranulation.}


   \maketitle


\section{Introduction}
Solar granulation is a manifestation of thermal convection; hot gas rises to the surface, cools, and sinks in the intergranular lanes. Granular structure is easily observed in white-light intensity images, even though the measured contrast is reduced by, e.g., stray light \citep[e.g.,][]{sanchez_2000}. After deconvolving the intensity images, the contrast is roughly 15\% RMS in the red \citep{wedemeyer_2009}.

The thermal signature of the larger-scale supergranulation \citep[see][for an extensive review]{rieutord_2010}, on the other hand, is hard to measure because
it is small compared to the granulation contrast 
and competes with the brightness increase due to the network magnetic field, which surrounds the supergranules \citep[e.g.,][]{liu_1974,foukal_1984}.
The latter effect results from reduced opacity in magnetic regions \citep[e.g.,][]{spruit_1976}.

Using observations from the ground-based PSPT, \citet{goldbaum_2009} and \citet{rast_2003} found a brightness excess of ${\sim}0.1\%$ in supergranules, corresponding to a temperature perturbation of ${\sim}1~$K. To obtain this result, the authors carefully removed magnetic pixels using \ion{Ca}{II}~K images, conducted ensemble averaging over thousands of supergranules and applied azimuthal averaging, thus losing spatial and temporal information.
 \citet{meunier_2007a} measured a similar (0.8$-$2.8~K) temperature excess using space-based SOHO/MDI intensity images and magnetograms (high-resolution mode) in combination with a magnetic field exclusion that takes into account neighboring pixels, but with a similar lack of spatial information.

Here we extent the previous studies on the supergranular brightness excess:
How does the convective intensity peak evolve? Does the intensity contrast show an east-west anisotropy, as the magnetic field \citep{langfellner_2015a} or wave travel times \citep{degrave_2015}? To tackle these questions, we make use of high-quality data from the Helioseismic and Magnetic Imager (HMI) \citep{schou_2012} onboard the SDO spacecraft.


\section{Observations and data processing}
We analyzed about one year of HMI data from May 2010 through May 2011, comprising Dopplergrams, continuum intensity maps, and line-of-sight magnetograms. These observables are computed from a combination of filtergrams that probe the \ion{Fe}{i} absorption line at 6173~\AA \ \citep{couvidat_2012a}.

For all three data products, we tracked the same regions of size ${\sim}180 \times 180~$Mm$^2$ at the solar equator for 40~hours. The tracking rate was chosen to match the rotation rate of the supergranulation pattern, which is $60$~m~s$^{-1}$ faster than the \citet{snodgrass_1984} rate at the equator \citep{gizon_2003}.
The cadence is 45~seconds; successive datacubes are spaced by 24~hours, resulting in 365 datacubes of size $40~\text{h} \times 180 \times 180~$Mm$^2$.
We remapped the regions using Postel's projection (we used this projection for all three data products) and a pixel size of $0.348~$Mm.

We divided the datacubes into five temporal segments of 8~hours length each; segment three crosses the central meridian. For each segment, we selected the f modes by applying the ridge filter described in \citet{langfellner_2015}. We then computed point-to-annulus wave travel-time differences \citep{duvall_1996}, which are sensitive to the horizontal divergence of the flow, with an annulus radius of 10~Mm, using the method of \citet{gizon_2004}. In these travel-time maps, we identified the locations of the minima and maxima, corresponding to the positions of strongest supergranular outflows and inflows in the horizontal plane \citep[see][for details of the identification method]{langfellner_2015}. For each segment, we identified roughly 8,000 outflow and inflow centers. Since the different segments cover the same solar regions and the supergranular lifetime is roughly one day, using all the five segments corresponds to an effective number of supergranules that is somewhat below the total number of 40,000 positions.

Using these coordinates, we co-aligned processed images of the continuum intensity contrast and the line-of-sight magnetic field, following the procedure described in \citet{langfellner_2015}.
The intensity datacubes were processed before the co-alignment in the following way. We divided the tracked and remapped datacubes into temporal segments in the same manner as the Doppler velocity datacubes described above. The individual segments were averaged over their respective length of 8~hours. Segments with notable activity (absolute value of intensity contrast $>0.1$, see below for definition) or an insufficient duty cycle (${<}90\%$ of frames) were excluded from further analysis. As the individual 8-hour datasets (for a given segment) contain independent realizations of supergranules, this exclusion of some datasets is not an issue. The main effect is a slightly increased noise level (compared to using all datasets) due to the variability of supergranulation, since there are fewer samples to average over. For our measurements, this increase of noise level is less than 14\% (in the worst case) compared to using all datasets.

We then fitted a fourth-degree two-dimensional polynomial (with $x$, pointing west, and $y$, pointing north in the Postel-projected maps, as the independent variables) to the average intensity images. This provided the large-scale intensity background, $I_0$, where spatial variations result from, e.g., limb darkening and unaccounted instrument systematics \citep{couvidat_2016}. For each pixel, we computed the intensity contrast, $\Delta I/I_0$, with $\Delta I = I-I_0$. Note that this approach also removes the effect of the instrument degradation, which led to a decrease in photon count of about 5\% over the total observation period \citep[cf.][]{cohen_2015}.

For the magnetic field, we averaged over 8~hours in the same fashion and then computed the absolute magnetic field to avoid cancellation of opposite polarities in the subsequent processing. This procedure is the same as in \citet{langfellner_2015a}. However, for the current study we have extended the observation period from four months to one year and studied five instead of three temporal segments of 8 hours. Further subtraction of the mean magnetic field map for each segment (averaged over all datacubes) yielded the deviation, $\Delta B$, from the background magnetic field strength. Contrary to the intensity, neither the mean nor the RMS of the line-of-sight magnetic field showed long-term trends over the observation period (except for increasing activity due to the solar cycle).


\section{Results}

\subsection{Spatially resolved supergranular intensity contrast}

   \begin{figure*}[h]
\sidecaption
\begin{minipage}[h][0.61\vsize][t]{0.7\hsize}
\begin{tabular}{cc}
\framebox{\large{Outflow}}  &  \framebox{\large{Inflow}}   \\
\includegraphics[width=0.46\hsize]{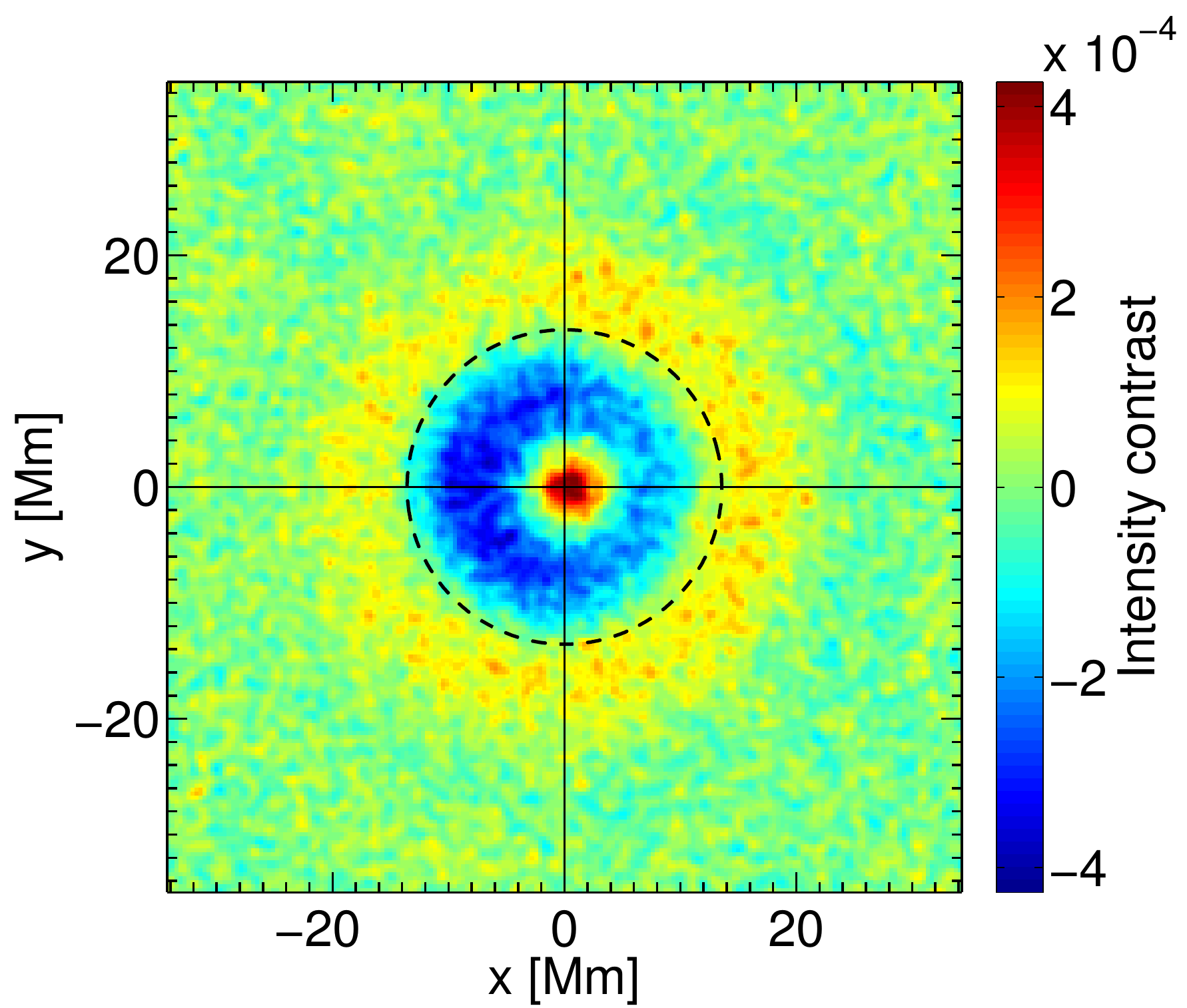} &
\includegraphics[width=0.46\hsize]{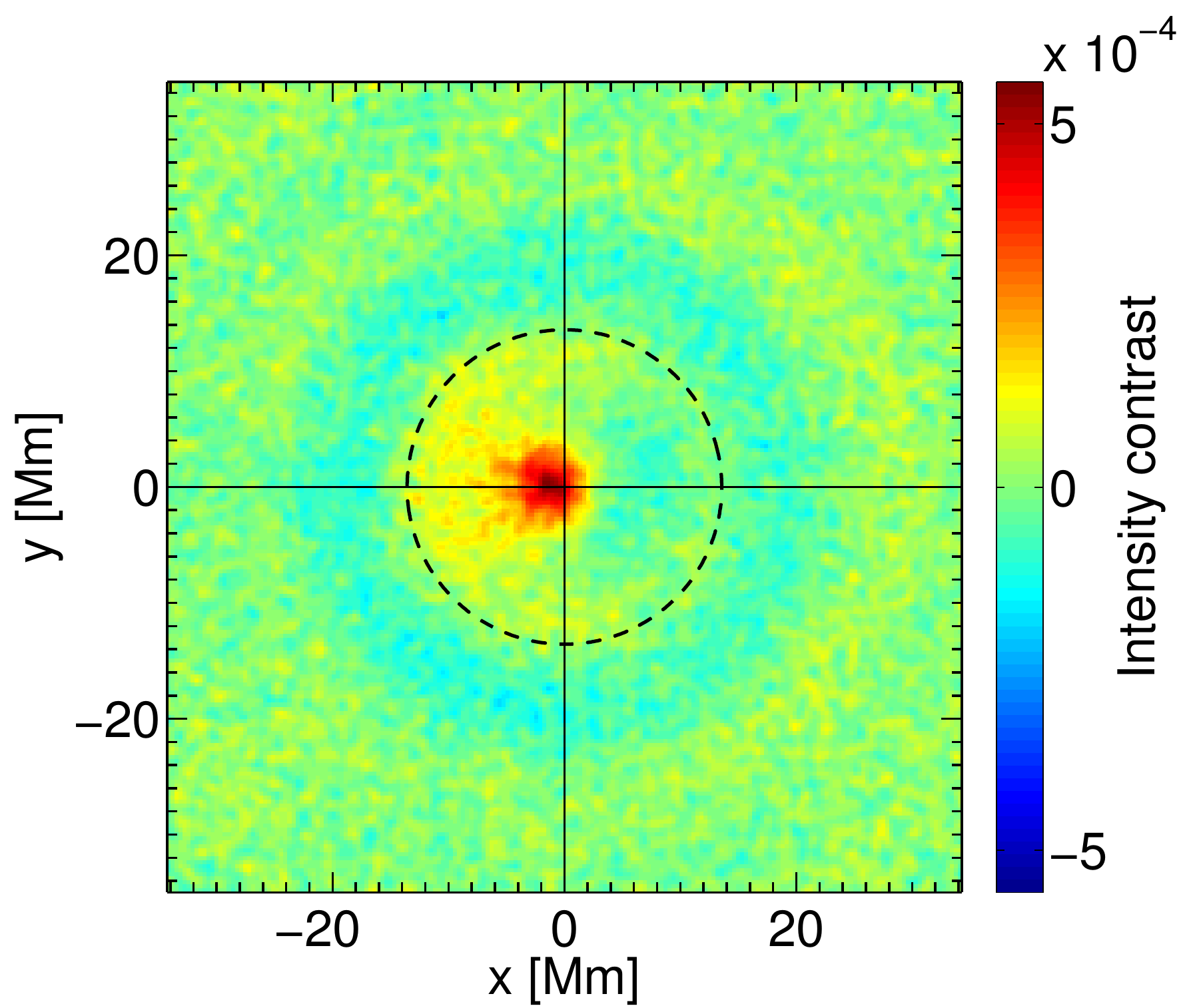} \\
\includegraphics[width=0.46\hsize]{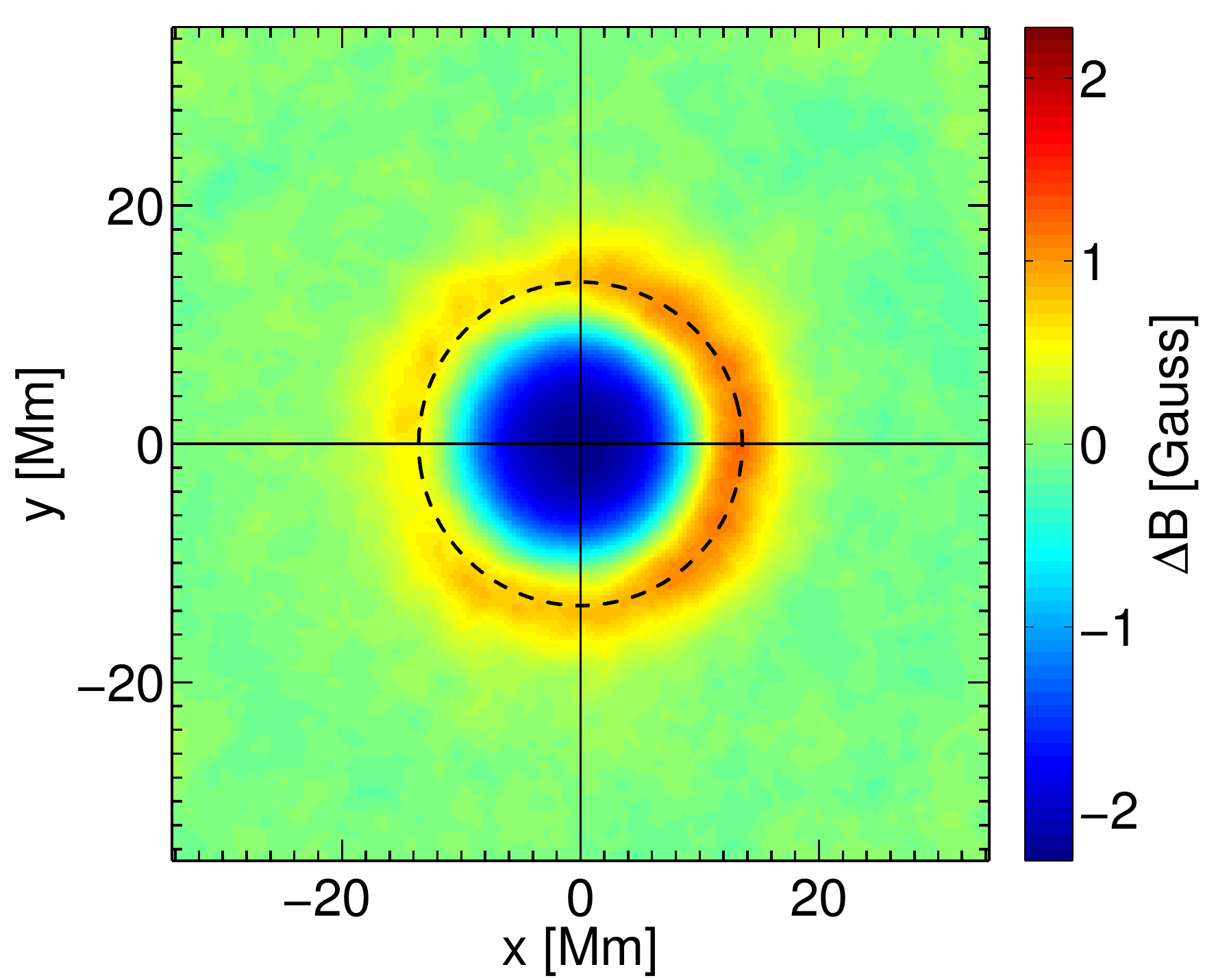} &
\includegraphics[width=0.46\hsize]{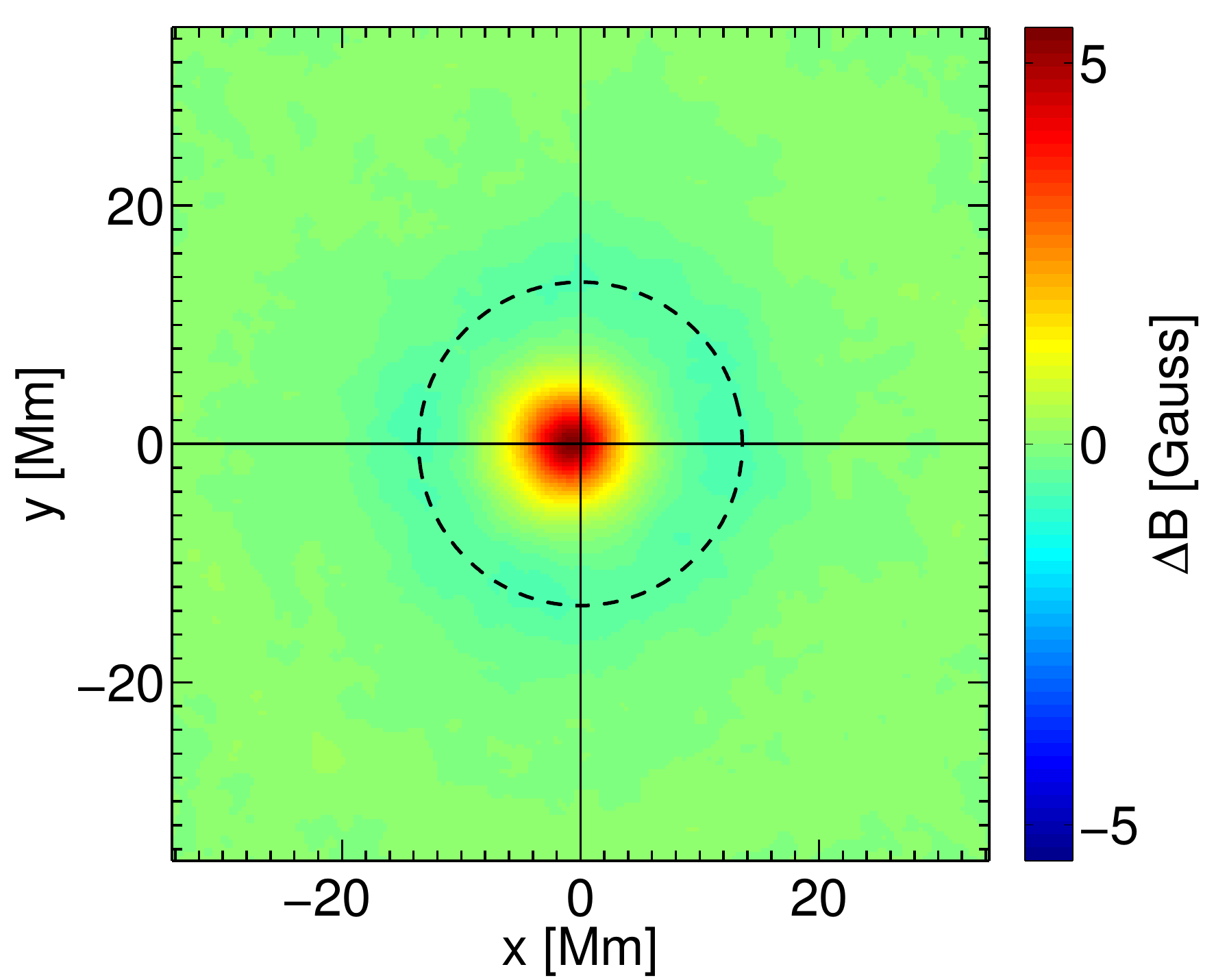}  \\
\includegraphics[width=0.495\hsize]{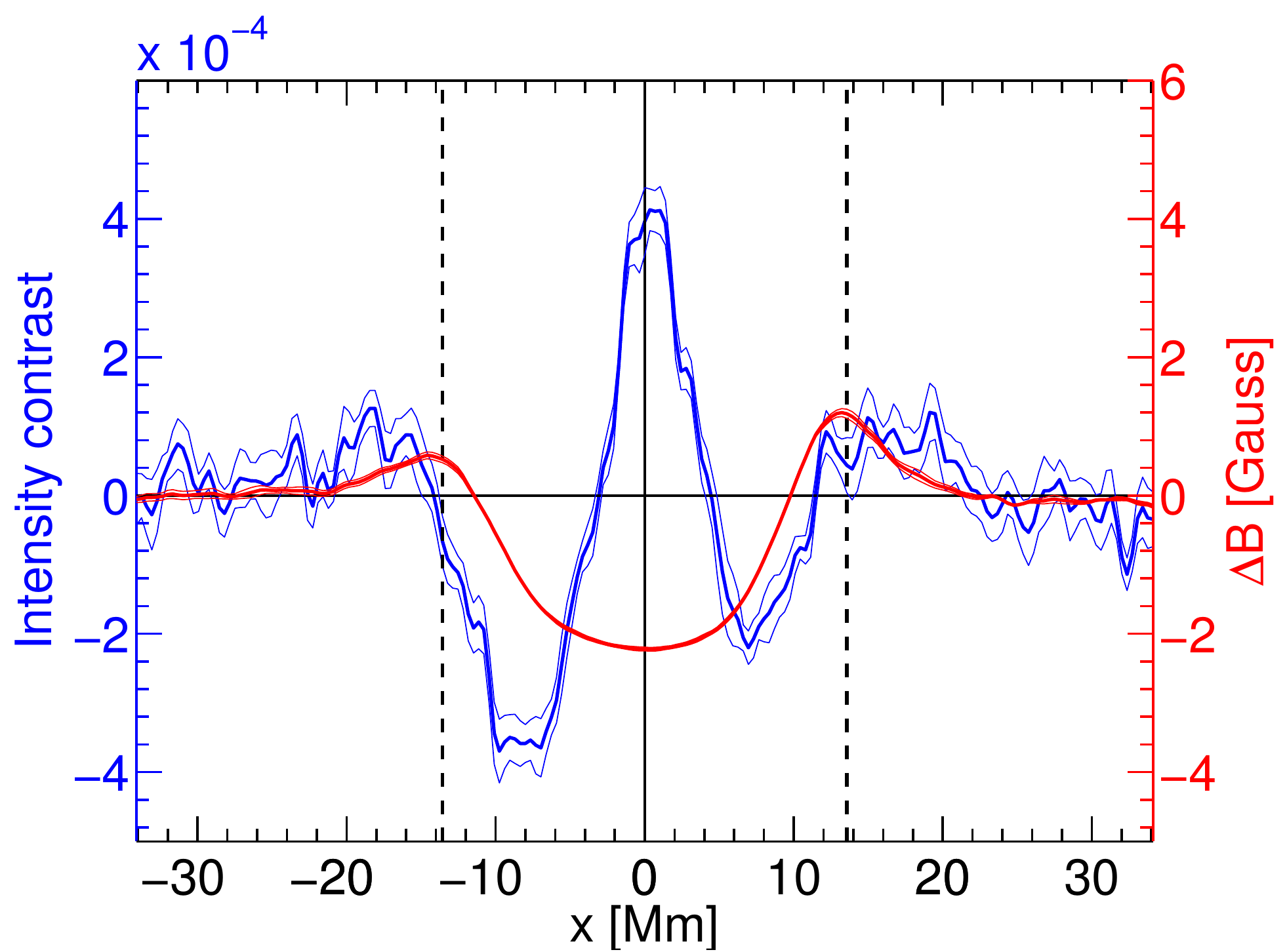} &
\includegraphics[width=0.495\hsize]{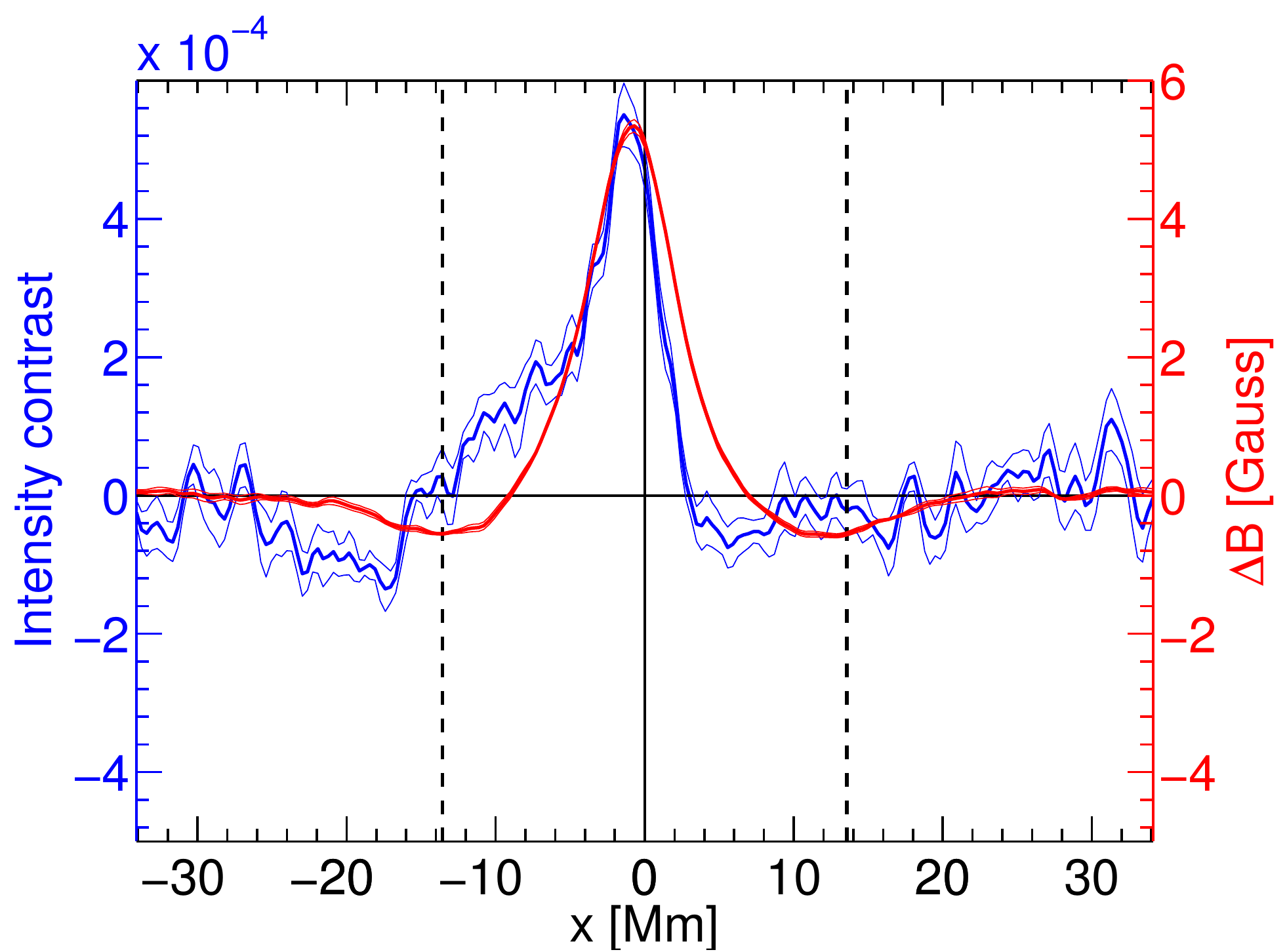}
\end{tabular}
\end{minipage}
\caption{Intensity contrast and magnetic field for the average supergranular outflow (\textit{left column}) and inflow (\textit{right column}): Mean over five 8~h segments at the equator around the central meridian, averaged over one year. \textit{Bottom row}: Cuts along $y=0$. The thin lines give the $1\sigma$ level of the variability, as computed from dividing one year of data into eight parts. The coordinates in all panels are given relative to the position of strongest horizontal outflow/inflow, measured with time-distance helioseismology. The $x$ coordinate points west, and $y$ is north. In all panels, the dashed circle has a radius of about 14~Mm.}
\label{fig1}
    \end{figure*}

The co-aligned average intensity contrast maps and magnetograms are shown in Fig.~\ref{fig1}. For the average supergranular outflow, the intensity contrast peaks at the center at about $4\times 10^{-4}$. This peak is surrounded by a ring of lower contrast at a radial distance of about 5 to 10~Mm, corresponding to the outer parts of the average supergranule. If measured from (east) minimum to maximum, the magnitude of the intensity peak is $(7.8\pm0.6)\times10^{-4}$ with a FWHM of 7$-$8~Mm. Further out, the intensity contrast increases above zero again, forming a second ring. This distance of 15 to 20~Mm is slightly larger than the distance where the surrounding inflows are located on average \citep[dashed line in Fig.~\ref{fig1}, or see][for more details]{langfellner_2015}.

The inner ring of lower contrast shows an east-west anisotropy; the dip is almost twice as deep in the east compared to the west. The significance of this anisotropy is greater than $6\sigma$ if an average over 10 pixels (3.5~Mm) is used for the comparison.
The outer ring of higher contrast appears to be stronger in the west, but without a high significance ($2.5\sigma$ if averaging over the east and west halves of the ring).

A comparison of the intensity contrast with the magnetic field strength at the same position reveals that the intensity peak occurs at the minimum of the broad magnetic field dip.
The minima of the intensity contrast roughly coincide with the inflection points of the magnetic field strength, and the outer zero crossings coincide with the maxima of the magnetic field and the surrounding inflows. The east-west anisotropy in the low-contrast region of the intensity has no obvious correspondence in the magnetic field. For the latter quantity, we recover the east-west anisotropy of the network field \citep{langfellner_2015a}, with a higher significance (${>}6\sigma$) than in the previous study (${>}3\sigma$).

For the average supergranular inflow, the intensity contrast shows a peak as well, with a magnitude ($5\times 10^{-4}$) that is comparable to the peak in the outflow. However, the peak is shifted eastward by about 1~Mm (corresponding to about three pixels), compared to the position of strongest inflow ($x=y=0$). At a radial distance of about 5 to 15~Mm, a strong east-west anisotropy is visible as an intensity bump in the east, and a dip in the west. The magnitude of the east-west difference is comparable with the anisotropy in the outflow. At about 20~Mm, the structure appears much more isotropic, in the form of a ring of lower intensity.

The magnetic field structure in the average inflow is much simpler. It features a strong peak (about 5~Gauss) that coincides with the east-shifted peak of the intensity. Further out, the magnetic field structure deviates strongly from the intensity structure. The peak is surrounded by a broad ring of weaker field between a radial distance of 10 and 20~Mm. This ring does not show any apparent anisotropy.

\subsection{Evolution of the intensity and the magnetic field}
To further characterize the intensity peak in the average supergranular outflow, we analyzed its temporal evolution. For this purpose, we used the supergranular outflow and inflow coordinates from a directly preceding or following temporal segment, yielding the intensity contrast and magnetic field both 8~hours before and after the reference time. This time span constitutes a significant fraction of the supergranular lifetime.

\begin{figure*}[h]
\sidecaption
\includegraphics[width=0.335\hsize]{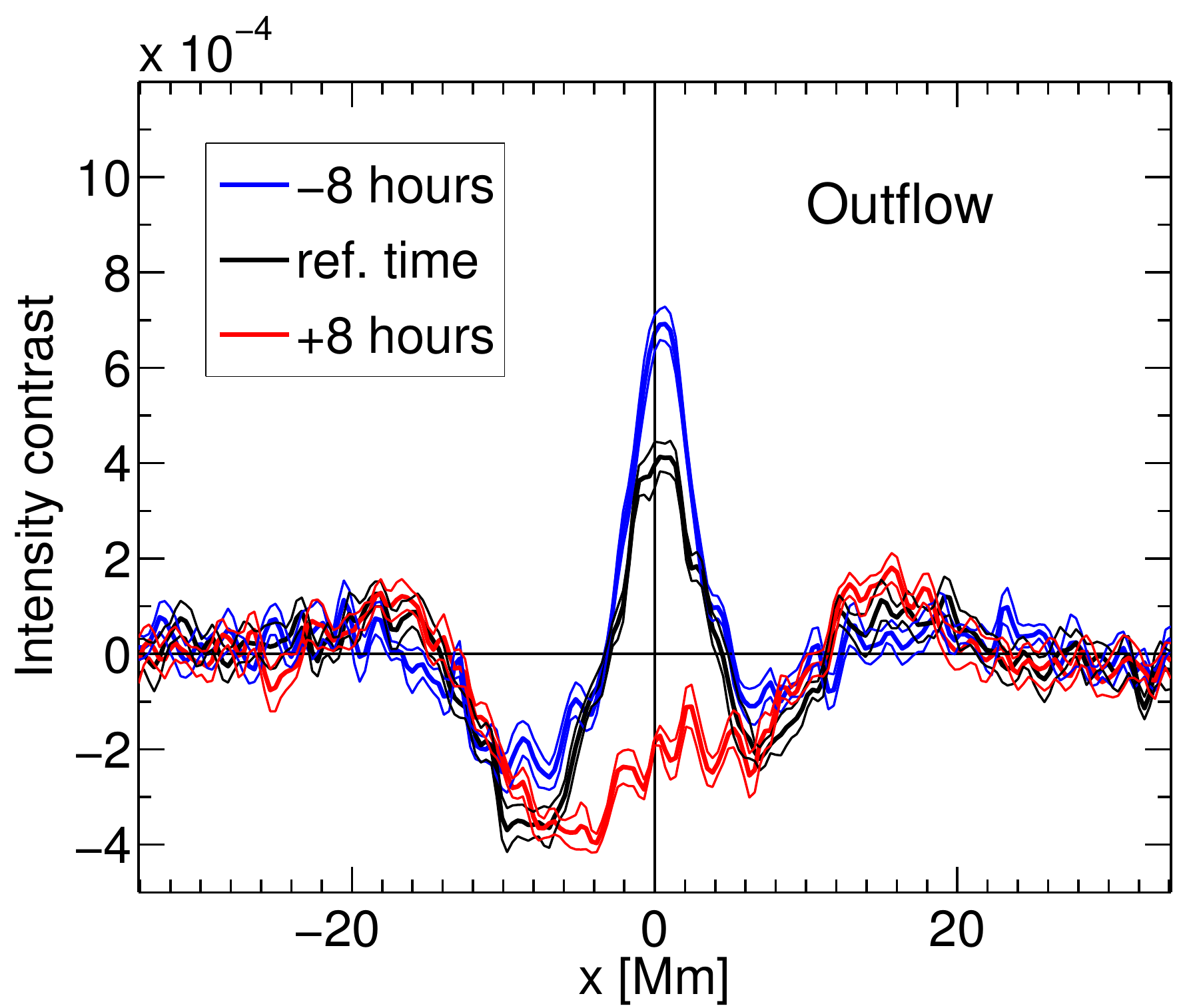}
\hspace{0.2cm}
\includegraphics[width=0.335\hsize]{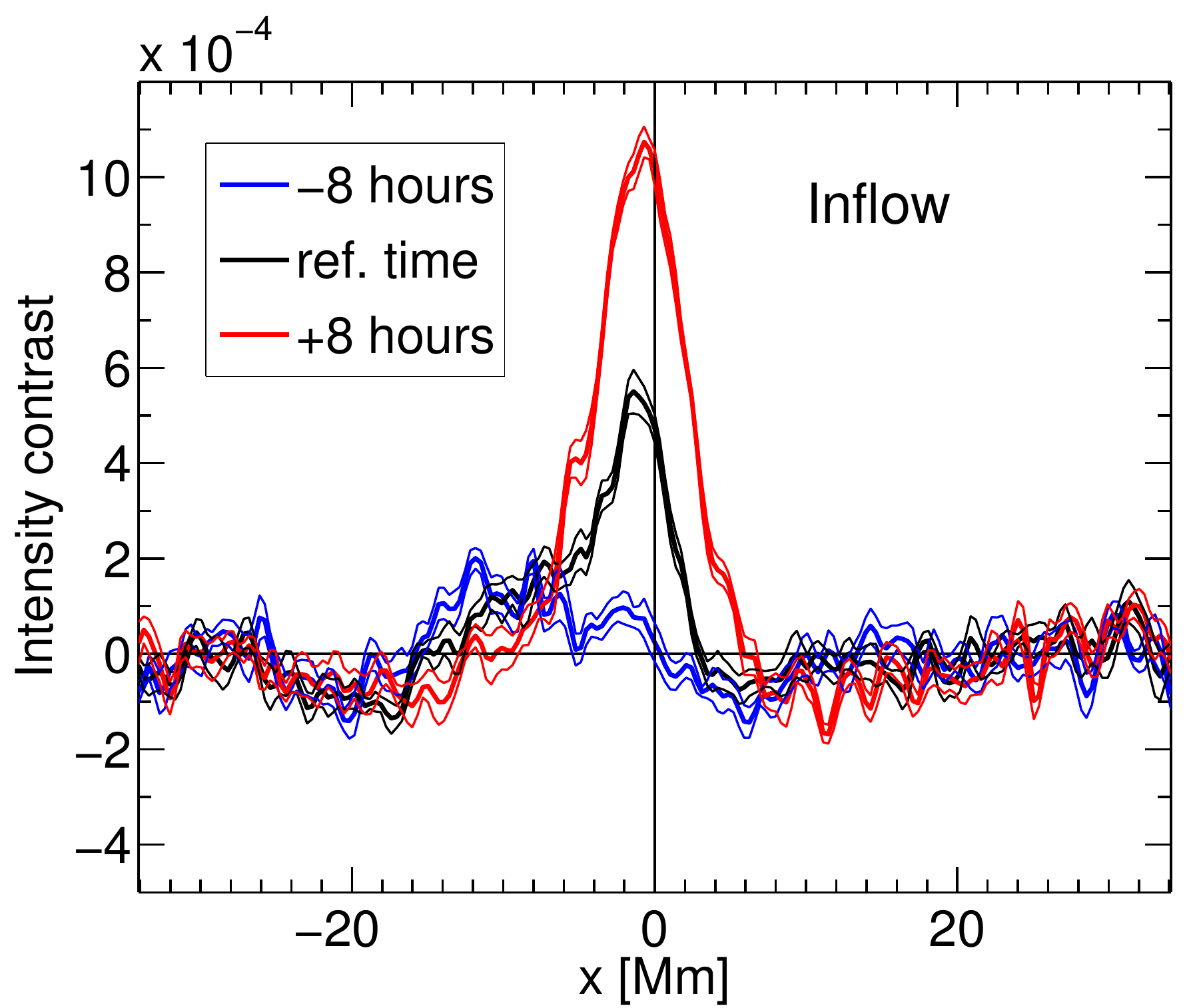}
\caption{Temporal evolution of the intensity contrast for the average supergranular outflow and inflow. Each curve is a mean over five 8~h segments at the equator around the central meridian, averaged over one year and cut along $y=0$. The thin lines give the $1\sigma$ level of the variability, as computed from dividing one year of data into eight parts.}
\label{fig2}
    \end{figure*}

As for the previous analysis, we averaged over five temporal segments for each datacube.
In order to provide all the needed supergranule coordinates in the case of the first and last segments, we extended the 40-hour datacubes by tracking for an additional 8~hours before and after. For these two new segments, we computed the locations of the supergranular outflows and inflows in the same way as for the other segments.
The results for averaging the intensity contrast over the five temporal segments are shown as cuts at $y=0$ in Fig.~\ref{fig2}. To facilitate the comparison, the results for the reference time are also plotted.

The evolution of the peak structures is dynamic and asymmetric. In general, we observe that the intensity and magnetic field structures differ strongly in the past but converge with time. The travel times (not shown), in contrast, evolve symmetrically in time, with the peak magnitudes decreasing by 25\% for the outflows, and 34\% for the inflows, both 8~hours before and after the reference time.

\begin{figure*}[h]
\sidecaption
\includegraphics[width=0.335\hsize]{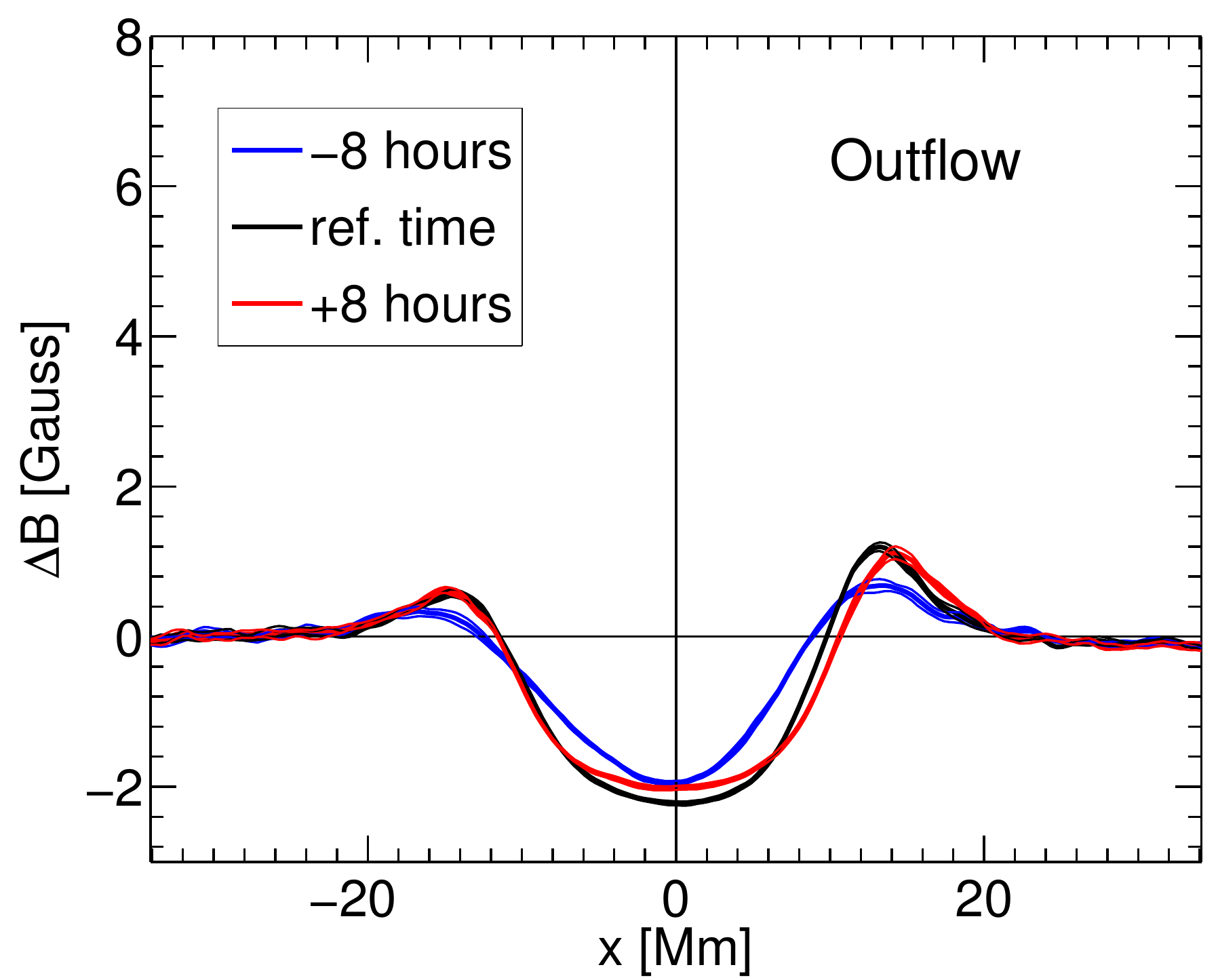}
\hspace{0.2cm}
\includegraphics[width=0.335\hsize]{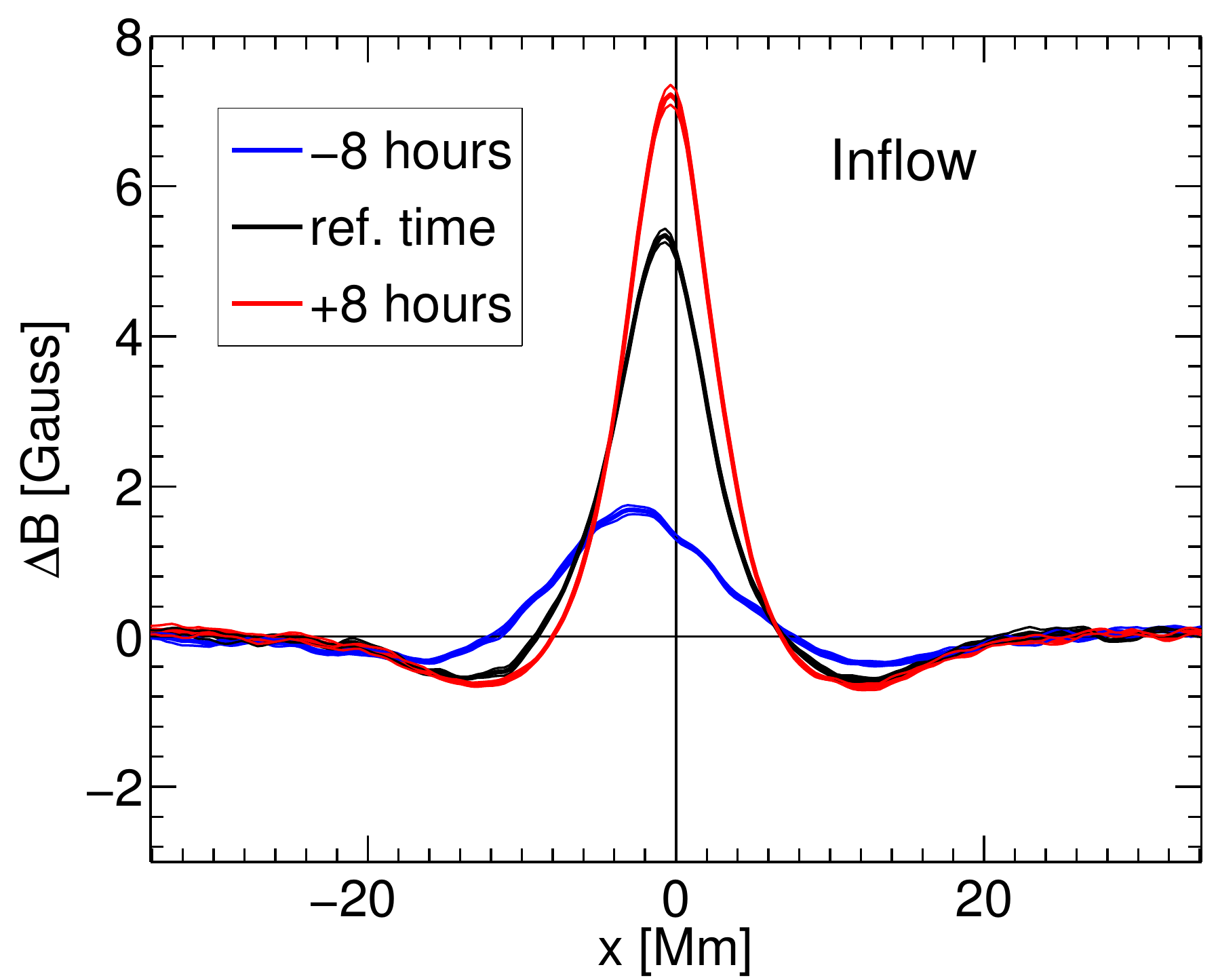}
\caption{Same as Fig.~\ref{fig2}, but for the magnetic field.}
\label{fig3}
    \end{figure*}

For the outflow, the intensity peak is strongest in the past (difference between minimum and maximum about $1\times 10^{-3}$), becomes slightly weaker at the reference time and has weakened dramatically 8~hours into the future. The peak intensity thus occurs before the strongest travel-time perturbation, which is a measure of the horizontal flow divergence. At the latter time, the overall structure has become similar to the magnetic field structure (see Fig.~\ref{fig3}), which has not changed except a slight broadening of the central dip. The magnetic field anisotropy (stronger field in the west) is now reflected in a similar anisotropy in the intensity. Repeating the significance test from the previous section yields a difference of more than $3\sigma$ between the west and east halves of the ring.

In the inflow, the evolution of the intensity peak shows the opposite behavior. In the past, there is no visible peak at the center, whereas 8~hours into the future it has grown to $1\times 10^{-3}$, which is twice the magnitude of the peak at the reference time. This development is accompanied by a qualitatively similar increase of the magnetic field strength peak. The east-west anisotropy in the intensity, on the other hand, is clearest in the past, whereas it has vanished in the future.

\subsection{Tests of robustness}  \label{sect_tests-robustness}
The measured anisotropies and evolution trends are robust features, which survived several tests (supplementary figures are provided in Appendix~\ref{sect_appendix}):

(i) Tracking at the equatorial \citet{snodgrass_1984} rate (instead of the supergranular pattern rotation rate) yielded the same results within error bars, except for a westward drift of the intensity and magnetic field structures (see Fig.~\ref{figA1}), corresponding to the difference in the tracking speed ($60~$m~s$^{-1}$, which is about 1.7~\text{Mm} per 8-hour timestep).

(ii) The intensity and magnetic field structures measured for the individual temporal segments (probing different longitudes up to ${\sim}{\pm}15\degr$ from central meridian) show only small variation (Fig.~\ref{figA2}), consistent with the error bars. Further toward the limb, an additional east-west anisotropy is present. This is likely caused by the Doppler signal of supergranular flows leaking into intensity due to imperfections in the HMI algorithm \citep[cf.~Fig.~5 in][]{cohen_2015}. However, this cross talk is removed by averaging over east and west of the central meridian, as we did.

We note that in the case of the magnetic field, the measured line-of-sight field strength is expected to decrease away from disk center if the field is vertical. However, even at 15$\degr$ away from disk center this decrease would be less than 4\%. As most of the observations are located much closer to the disk center and since the curves in Fig.~\ref{figA2} show no such trend, we did not apply any correction with respect to the field orientation.

(iii) Using randomized coordinates (instead of supergranule positions) yielded maps without any visible structures except for a flat noise background (Fig.~\ref{figA3}).


\section{Discussion and conclusions}
Using Planck's law, we have $\Delta I/I_0 \approx 4\Delta T/T_0$, where $T_0=5777~$K corresponds to HMI's iron absorption line (6173~\AA). With $\Delta I/I_0 = (7.8\pm0.6)\times10^{-4}$ at the center of the outflow, this gives $\Delta T \approx 1.1\pm0.1$~K. This result is consistent with the value ${\sim} 1$~K obtained by \citet{goldbaum_2009} (no error bar provided) and the range 0.8$-$2.8~K measured by \citet{meunier_2007a}.

The east-west anisotropy of the intensity contrast that we detect at the equator consists of two components: (i) an anisotropy in the network that has the same spatial pattern as the already known magnetic field anisotropy \citep{langfellner_2015a} and (ii) an anisotropy that is distinct from the magnetic field anisotropy, presumably of convective origin. Since the opacity is reduced in magnetic regions and the brightness is increased, the anisotropy of type (i) can be regarded as an independent confirmation of the magnetic field anisotropy, using a different observable.

The east-west anisotropies of the intensity and magnetic field signals at the equator are most likely connected to the travelling-wave properties of supergranulation and the superrotation of the pattern \citep{gizon_2003,schou_2003}. This connection remains to be specified by studying the evolution of the intensity and the magnetic field signals over longer times (several days).

The decrease of the intensity peak from the time of maximum divergence can be interpreted as the termination of the driving of the supergranular outflow. The temperature excess may imply a pressure excess in the supergranule center that would accelerate the plasma horizontally. Due to inertia, the outflow does not stop immediately, but continues and weakens over time.

After the convective brightness excess has vanished, the intensity contrast reflects the magnetic field distribution. The broadening of the magnetic field dip in the outflow region and the accumulation of magnetic field in the network beyond the time of maximum outflow, are qualitatively consistent with the assumption that supergranular flows advect the magnetic field \citep{orozco_2012}.

Alternative models have been proposed that treat supergranulation not as a convective phenomenon but as a pattern resulting from the collective (non-linear) interaction of granules \citep{rieutord_2000,rast_2003a} or magnetic elements \citep{crouch_2007}. For example, the model by \citeauthor{crouch_2007} implies that the network magnetic field builds up before the supergranular inflows, contrary to what our measurements indicate. These alternative models do not make clear predictions for the intensity contrast, but our findings may impose additional constraints that could be incorporated in the future.

\begin{acknowledgements}
The HMI data used are courtesy of NASA/SDO and the HMI science team.
The data were processed at the German Data Center for SDO (GDC-SDO), funded by the German Aerospace Center (DLR). Support is acknowledged from the SPACEINN and SOLARNET projects of the European Union.
We thank B.~L\"optien for providing comparative data on the HMI Doppler/intensity cross-talk and J.~Schou, R.~Cameron, T.~Duvall, and B.~Beeck for the useful discussions and helpful comments on the manuscript. We are grateful to R.~Burston and H.~Schunker for providing help with the data processing, especially the tracking and mapping. We used the workflow management system Pegasus funded by The National Science Foundation under OCI SI2-SSI program grant \#1148515 and the OCI SDCI program grant \#0722019.
\end{acknowledgements}

\bibliographystyle{aa}
\bibliography{literature}

\appendix

\section{Supplementary figures}   \label{sect_appendix}

\newpage

   \begin{figure}
\includegraphics[width=0.8\hsize]{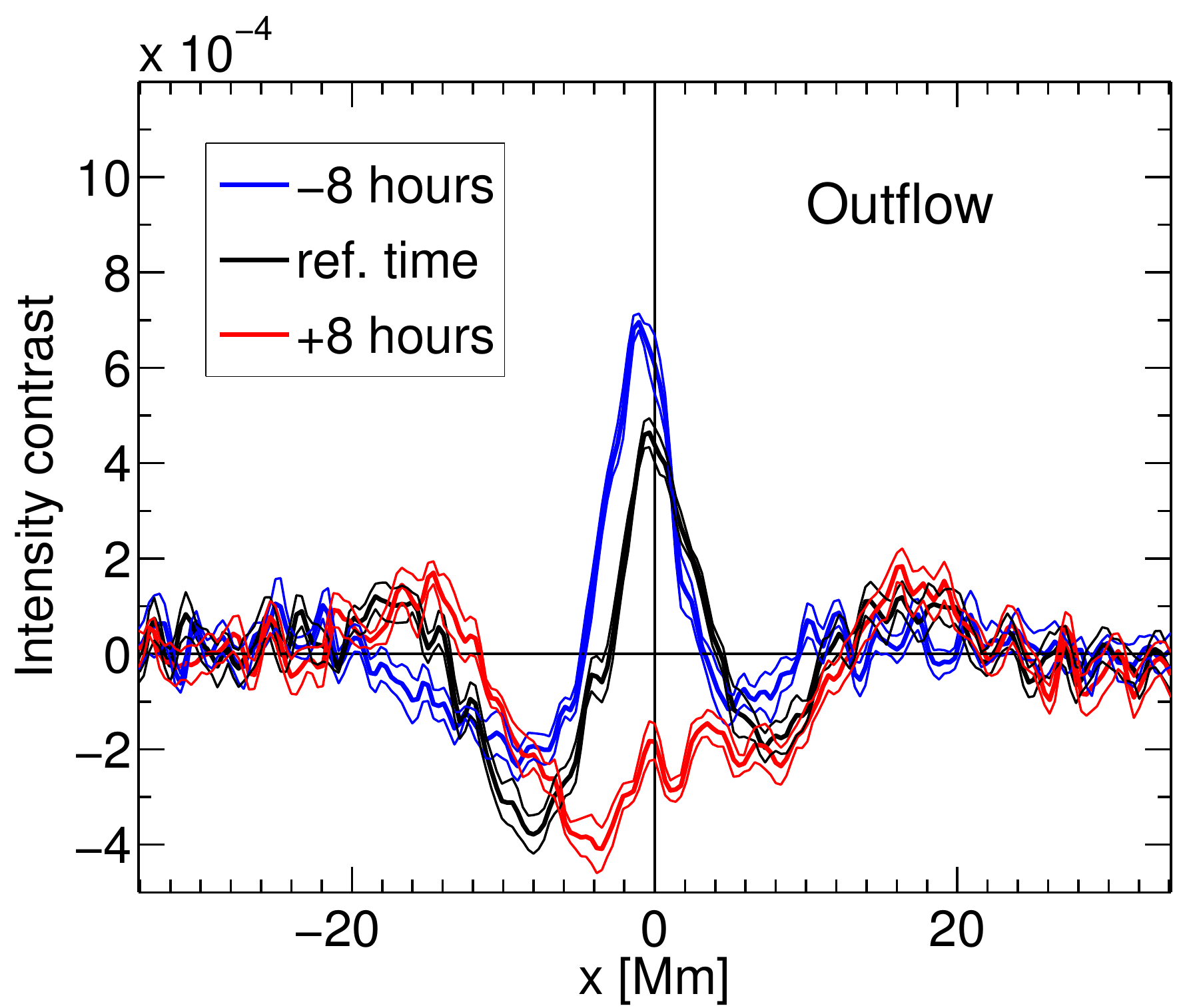}
\hspace{0.2cm}
\includegraphics[width=0.8\hsize]{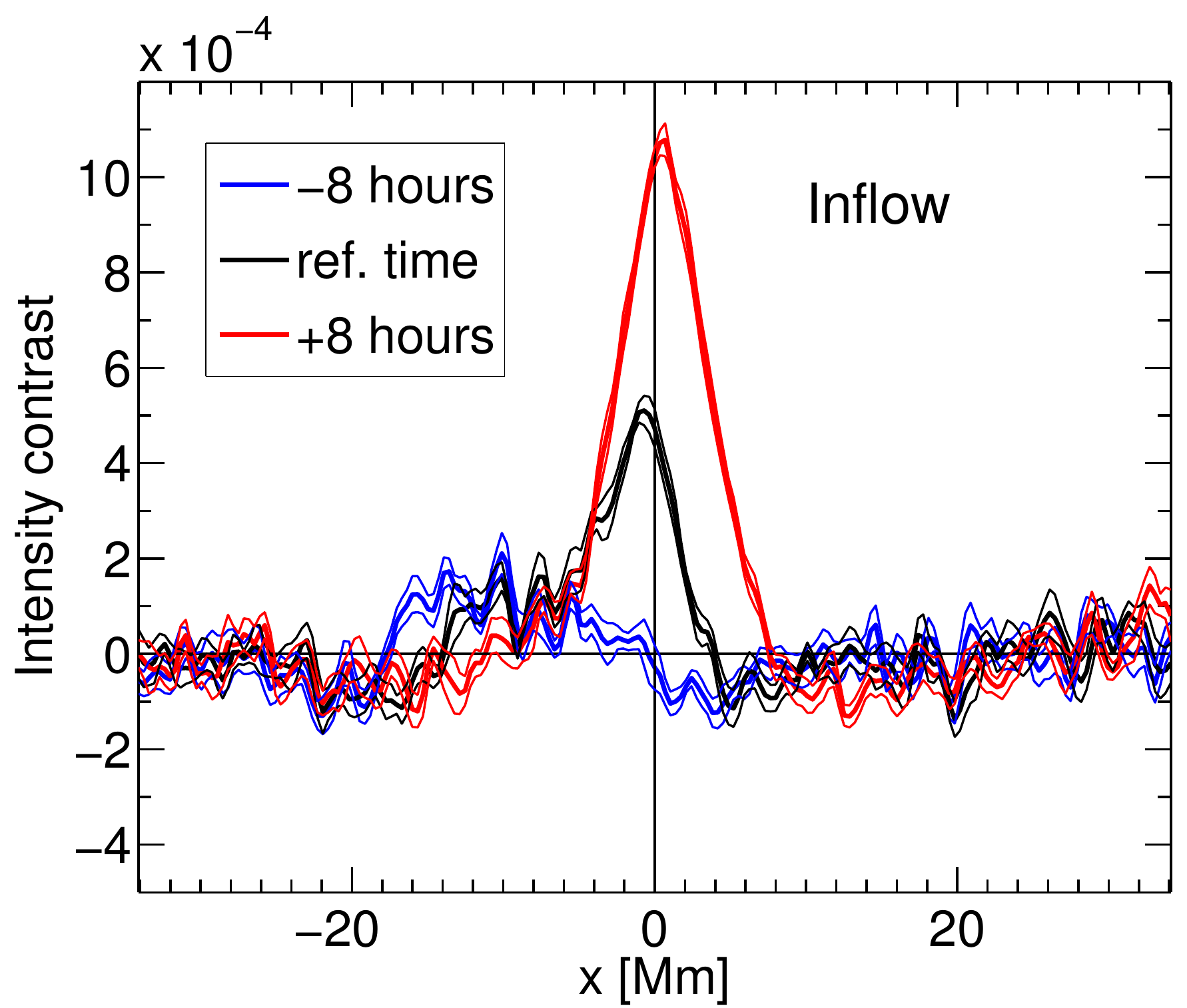}
\caption{As Fig.~\ref{fig2}, but for a tracking rate that is slower by $60~$m~s$^{-1}$ (see Sect.~\ref{sect_tests-robustness}).}
\label{figA1}
    \end{figure}

   \begin{figure*}
\centering
\includegraphics[width=0.4\hsize]{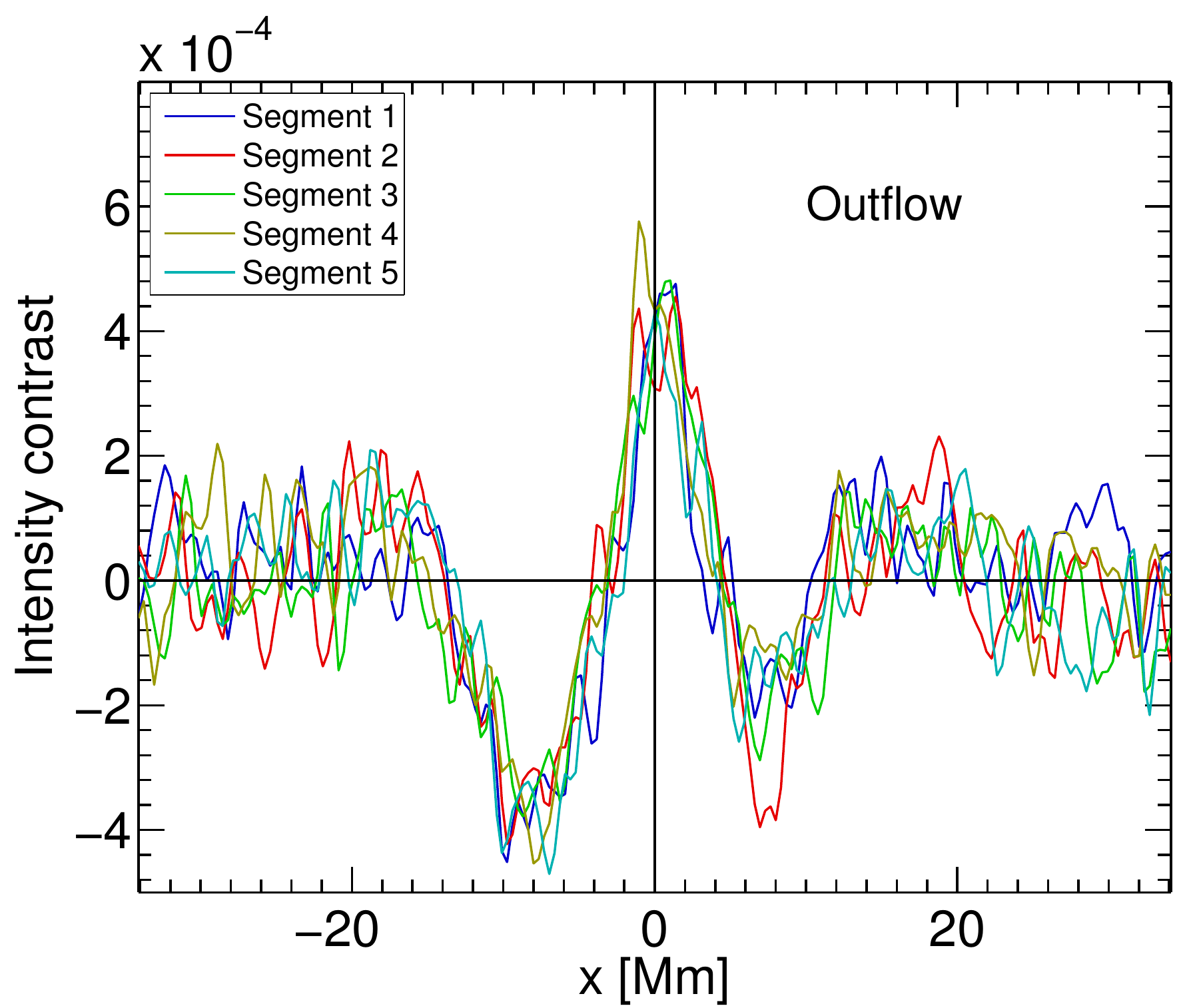}
\hspace{0.2cm}
\includegraphics[width=0.4\hsize]{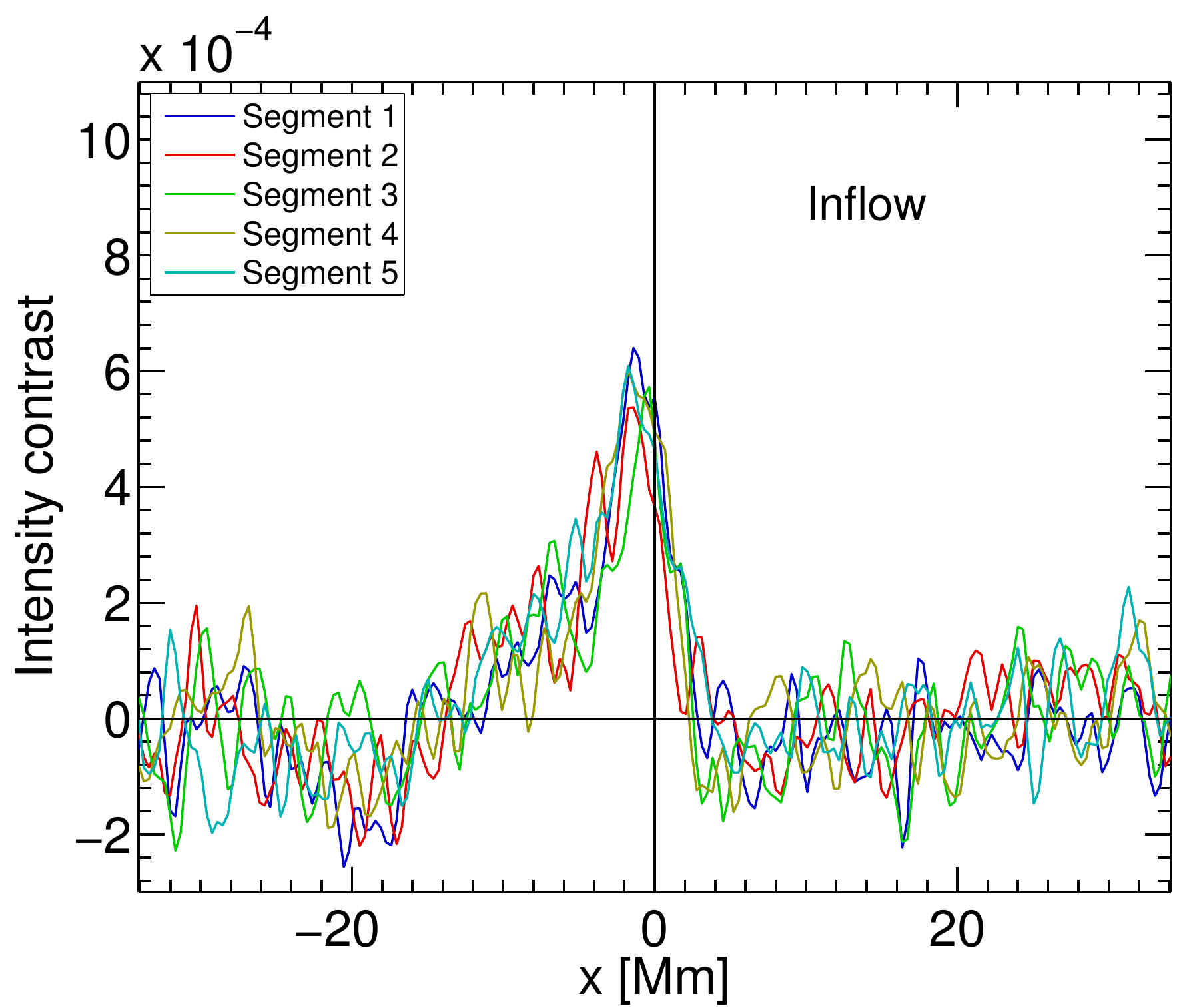}
\\
\includegraphics[width=0.4\hsize]{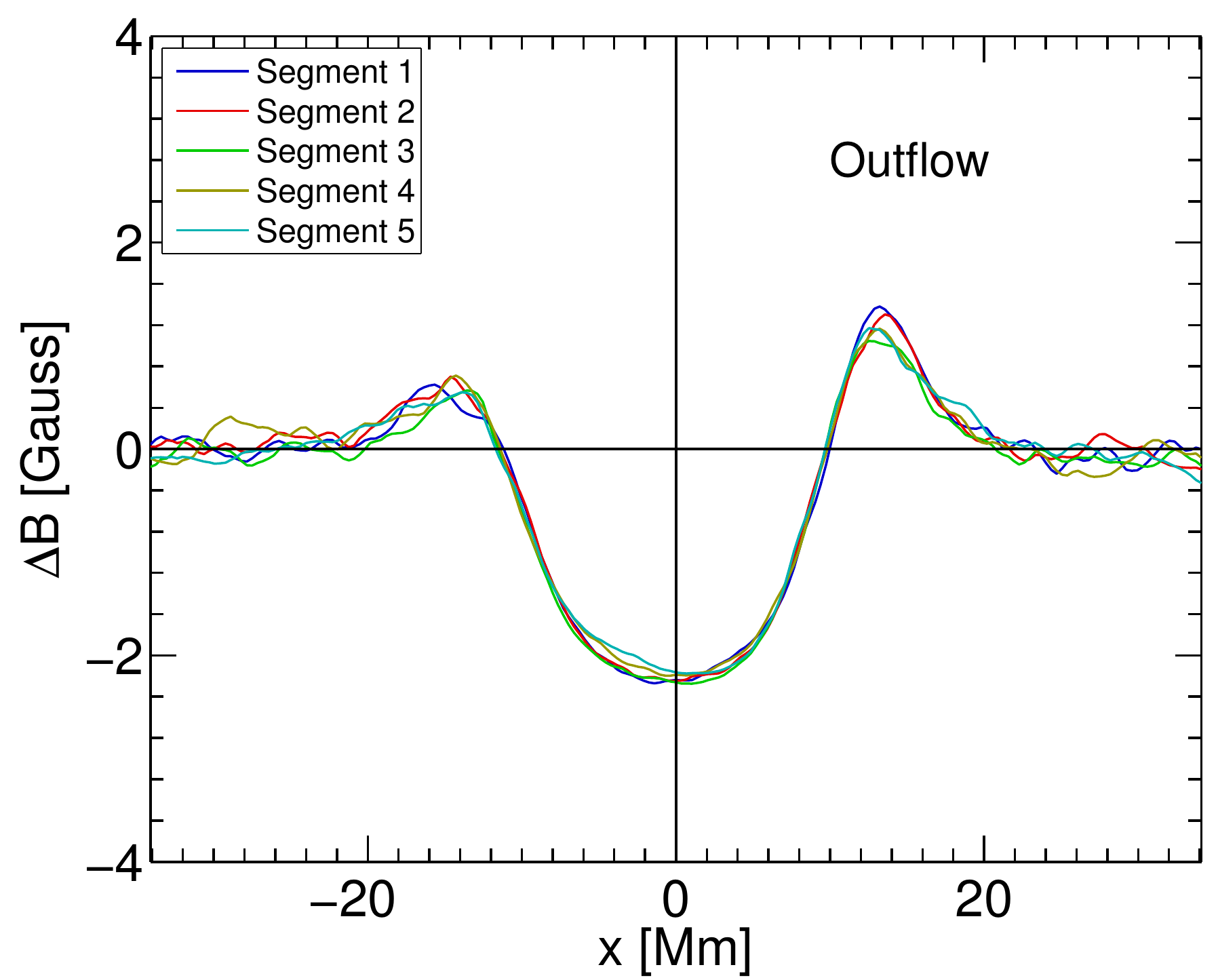}
\hspace{0.2cm}
\includegraphics[width=0.4\hsize]{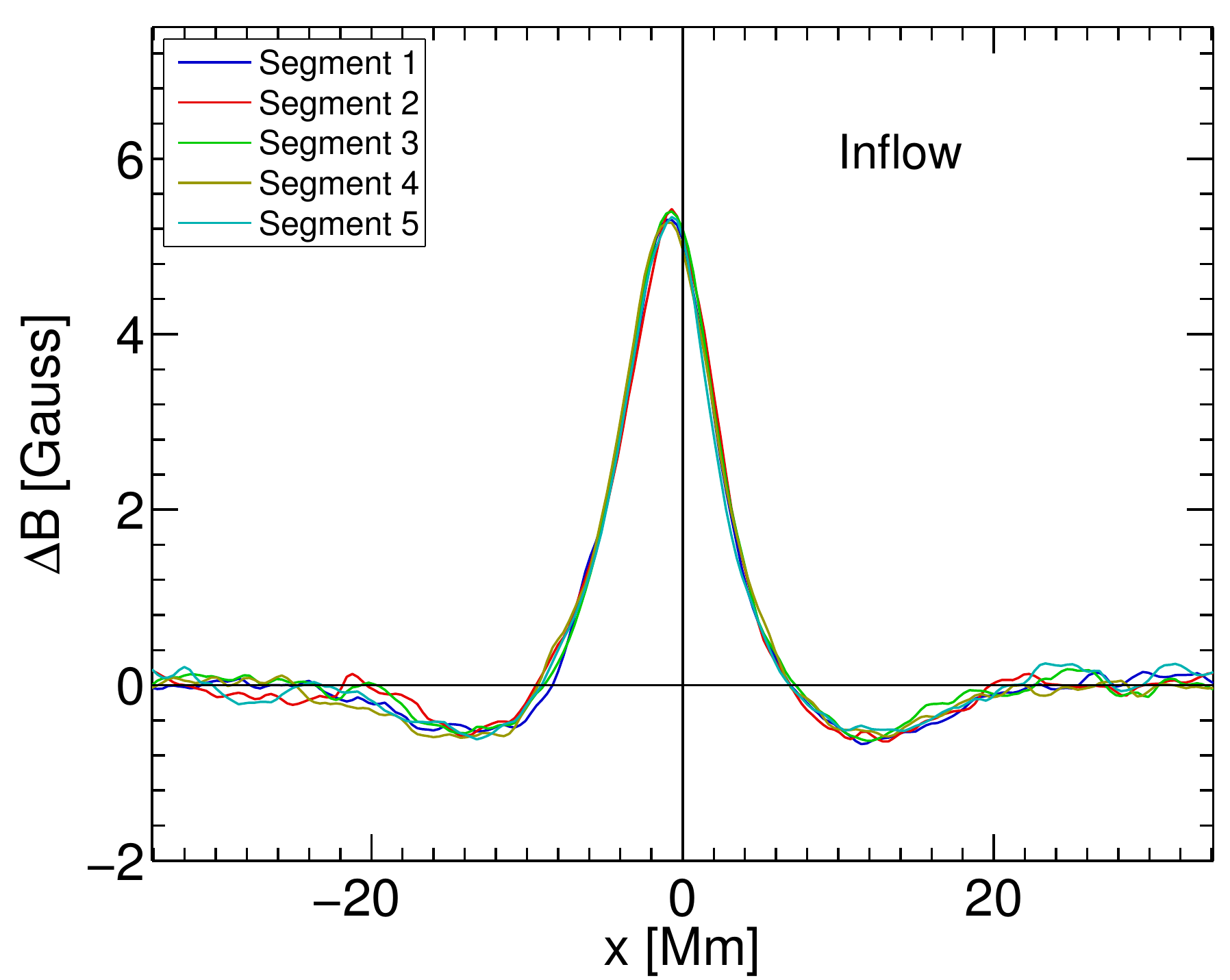}
\caption{Quantities as in the bottom row of Fig.~\ref{fig1}, but for single temporal segments. Segment~3 crosses the central meridian. For clarity, the intensity contrast and magnetic field are shown in separate panels.}
\label{figA2}
    \end{figure*}

   \begin{figure}
\centering
\includegraphics[width=0.9\hsize]{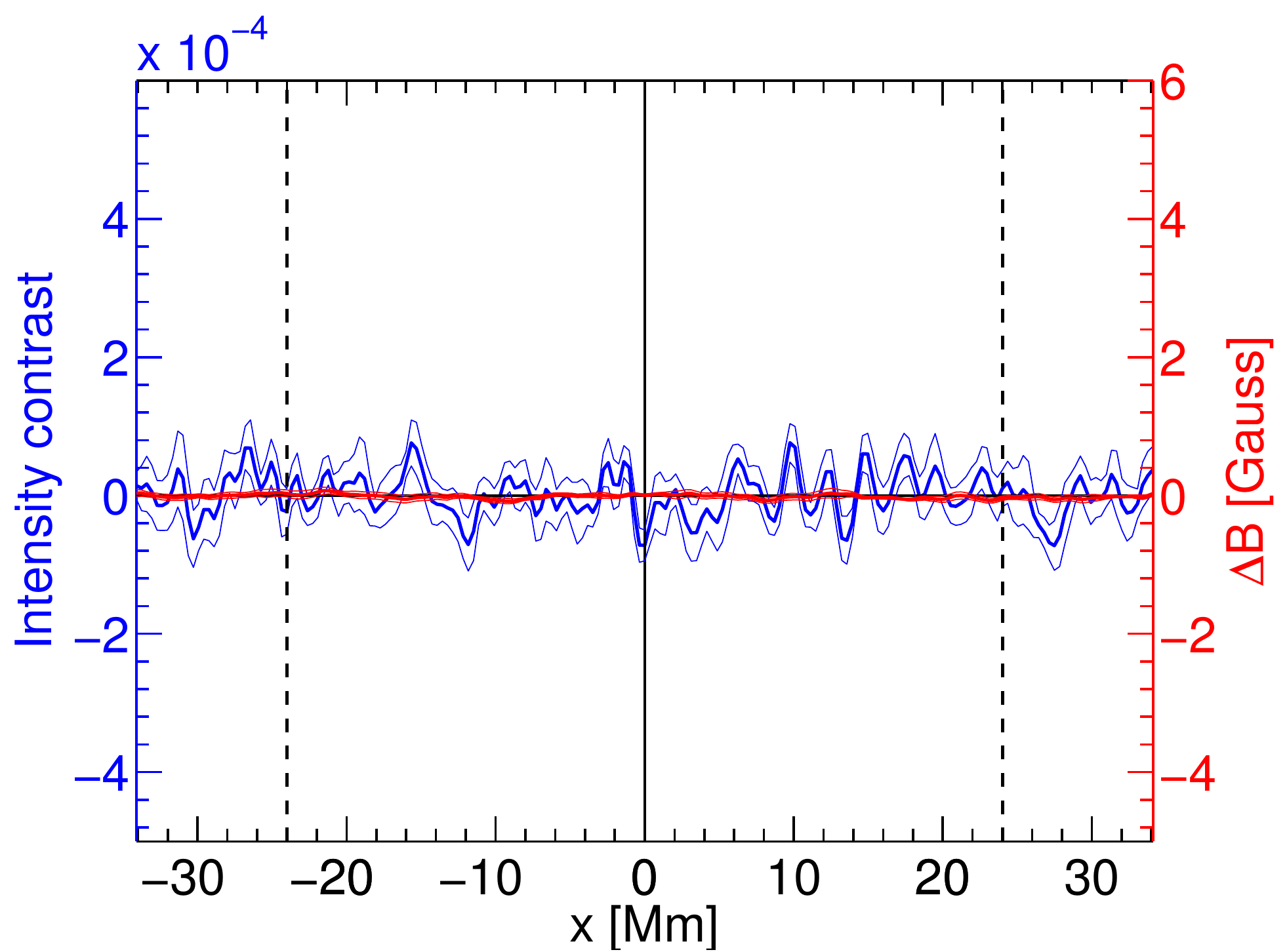}
\caption{As bottom row of Fig.~\ref{fig1}, but using randomized coordinates instead of supergranule positions.}
\label{figA3}
    \end{figure}

\end{document}